\documentstyle[12pt,epsfig]{article}
\voffset     = - 0.10 in
\begin{document}
\thispagestyle{empty}

{\Large\bf
\begin{center}
%Experimental Study of 
Effect on a Hadron Shower Leakage on the Energy Response and 
Resolution of Hadron TILE Calorimeter 
\end{center}
}

%\bigskip

\begin{center}
{\large\bf
Y.A.~Kulchitsky, V.S.~Rumyantsev \\
}
{\it
Institute of Physics, Academy of Sciences, Minsk, Belarus \\
\& JINR, Dubna, Russia
}

{\large\bf
J.A.~Budagov, V.B.~Vinogradov \\
}
{\it
JINR, Dubna, Russia
}

{\large\bf
G.~Karapetian, M.~Nessi \\
}
{\it
CERN, Geneva, Switzerland
}

{\large\bf
A.A.~Bogush \\
}
{\it
Institute of Physics, Academy of Sciences,  Minsk, Belarus
}
\end{center}

\begin{abstract}
The hadronic shower longitudinal and lateral leakages and its effect
on the pion response and energy resolution of $ATLAS$
iron-scintillator barrel hadron prototype calorimeter with
longitudinal tile configuration with a thickness of $9.4$ nuclear
interaction lengths have been investigated.
The results are based on $100\ GeV$ pion beam data at incidence angle
$\Theta = 10^{o}$ at impact point $Z$ in the range from $- 36$ to $20\ cm$
which were obtained during test beam period in May 1995
with setup equipped scintillator detector planes placed behind and back
of the calorimeter.
The fraction of the energy of $100\ GeV$ pions at $\Theta = 10^{o}$
leaking out at the back of this calorimeter amounts to $1.8 \%$
and agrees with the one for a conventional 
iron-scintillator calorimeter. 
Unexpected behaviour of the energy resolution as a function of leakage is
observed:
$6 \%$ lateral leakage lead to $18 \%$ improving of energy resolution in
compare with the showers without leakage.
The measured values of longitudinal 
punchthrough probability 
$(18 \pm 1) \%$ and $(20 \pm 1) \%$
for two different hit definitions of leaking events 
agree with the earlier  measurement for our
calorimeter and with the one for a conventional iron-scintillator calorimeter
with the same nuclear interaction length thickness respectively.
Due to more soft $cut$ for hit definition in the leakage detectors
the measured value of longitudinal punchthrough probability more 
corresponds to 
the calculated iron equivalent length $L_{Fe} = 158\ cm$.
\end{abstract}

\newpage

\begin{figure*}[tbph]
     \begin{center}
       %\vspace*{1in}
      \mbox{\epsfig{figure=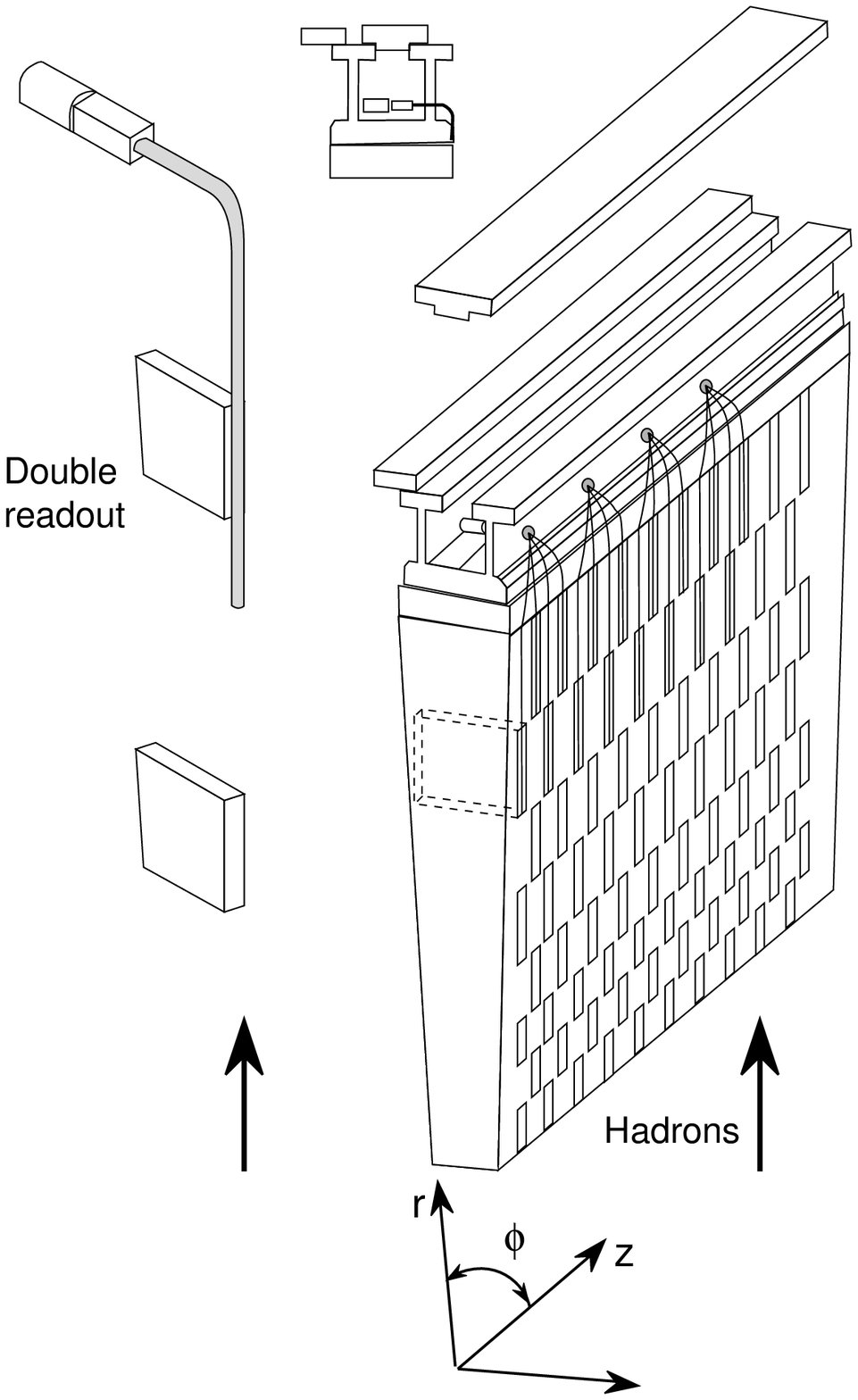,width=0.85\textwidth,height=0.9\textheight}}
     \end{center}
       \caption{
       Principal of the tile hadronic calorimeter.
       \label{fig:f1}}
\end{figure*}

\begin{figure*}[tbph]
     \begin{center}
        \begin{tabular}{c}
        %\hline
      \mbox{\epsfig{figure=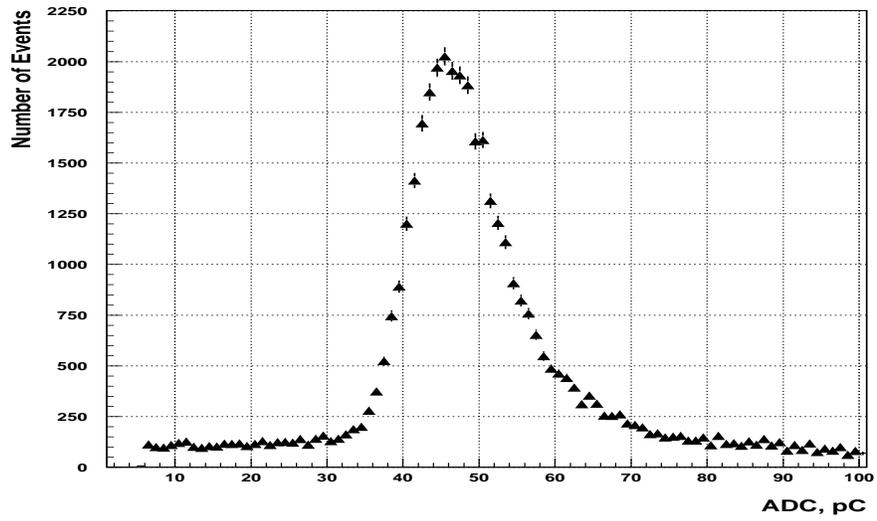,width=0.95\textwidth,height=0.4\textheight}} \\
        %\hline
        %\hline
      \mbox{\epsfig{figure=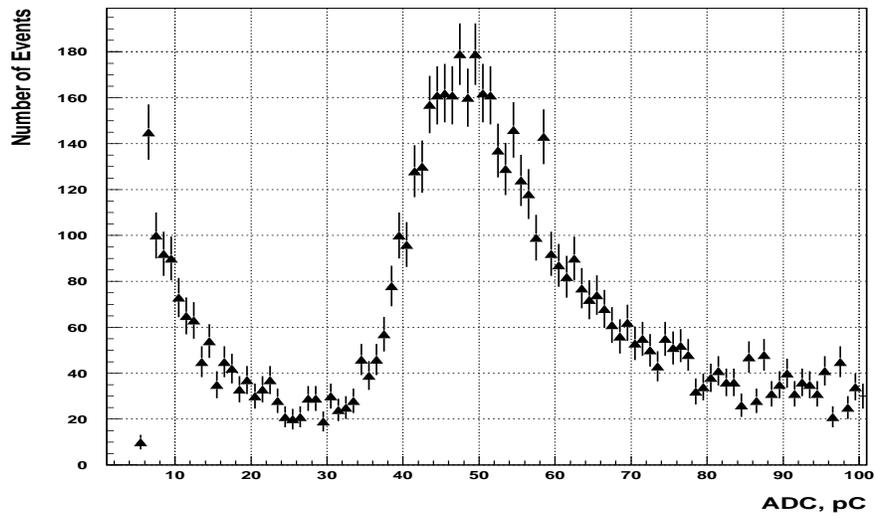,width=0.95\textwidth,height=0.4\textheight}} \\
        %\hline
        \end{tabular}
     \end{center}
      \caption{
        Typical $ADC$ spectrum of a ``{\it muon wall}''
         counters in the $\mu$ beam (top)
        and in the $\pi$ beam (bottom)
        for counter $N^{\b{o}}\ 8$.
        \label{fig:f4}}
\end{figure*}

{
\section{Introduction}

Due to limited dimensions of calorimeters one from important questions
of calorimetry concerns the energy leakage and related with it the
deterioration of energy resolution, appearance of tails in the energy
distributions and ultimately the deterioration of the quality obtained
physics information.
In this article we report on the results of the experimental study
of hadronic shower leakage effects on the pion response and
energy resolution of $ATLAS$
barrel hadron prototype calorimeter \cite{atcol}.
Because this calorimeter has innovative concept of longitudinal
segmentation of active and passive layers (see
Fig.~\ref{fig:f1}),
the measurement of hadron showers leakage is of special interest 
\cite{gild-91}.
This investigation was performed on the basis of data from $100\ GeV$ pion
exposure of the prototype calorimeter
at the $H8$ beam of the $CERN$ $SPS$
at different $Z$ impact points
in the range from $- 36$ to $20\ cm$ with step $2\ cm$
($Z$ scan) at incident angle $\Theta = 10^{o}$ which were obtained
in May 1995.
Earlier some results related with leakage for this calorimeter were
obtained in \cite{ber}, \cite{bos}, \cite{loc64}.

\section{The Prototype Calorimeter}

The prototype calorimeter is composed of five sector modules, each
spanning $2 \pi / 64$ in azimuth, $100\ cm$ in the axial ($Z$) direction,
$180\ cm$ in the radial direction,
and with  a front face of $100 \times 20\ cm^2$ \cite{ber}.
The iron structure of each module consists of 57 repeated ``periods''.
Each period is $18\ mm$ thick and consists of four layers.
The first and third layers are formed by large trapezoidal steel plates
(master plates), $5\ mm$ thick and spanning the full radial dimension of the
module.
In the second and fourth layers, smaller trapezoidal steel plates
(spacer plates) and scintillator tiles alternate along the radial direction.
The spacer plates are $4\ mm$ thick and of 11 different sizes.
Scintillator tiles are $3\ mm$ thickness.
The iron to scintillator ratio is 4.67:1 by volume.
The calorimeter thickness at incidence angle $\Theta = 10^{o}$ 
corresponds to $158\ cm$ of iron equivalent
(9.4 nuclear interaction length) \cite{loc64}.

Radially oriented $WLS$ fibres collect light from the tiles at both of
their open edges and bring it to photo-multipliers ($PMTs$) at the periphery
of the calorimeter.
Each $PMT$ views a specific group of tiles, through the corresponding
bundle of fibres. 
With this readout scheme three-dimensional segmentation
is immediately obtained.

Tiles of 18 different shapes all have the same radial dimen\-si\-ons 
($10\ cm$).
The prototype calorimeter is radially segmented into four depth segments by
grouping fibres from different tiles.
Proceeding outward in radius, the three smallest tiles,
$1  \div  3$ are grouped into section $1$,
$4  \div  7$ into section $2$,
$8  \div 12$ into section $3$ and
$13 \div 18$ into section $4$.
The readout cell width in $Z$  direction is about $20\ cm$.

Construction and performance of $ATLAS$
iron-scintillator barrel had\-ron prototype calorimeter is described
elsewhere \cite{atcol}, \cite{ber}, \cite{bud}.

\section{Test Beam Layout}

The five modules have been positioned on a scanning table, able to allow
high precision movements along any direction.
Upstream of the calorimeter, a trigger counter telescope was installed,
defining a beam spot of $2\ cm$ diameter.
Two delay-line wire chambers, each with $Z$, $Y$  readout,
allowed to reconstruct the impact point of beam particles on the
calorimeter face to better than $\pm 1\ mm$ \cite{ariz-94}.
For the detection of the hadronic shower longitudinal and lateral leakages
backward ($80 \times 80\ cm^2$) and side ($40 \times 115\ cm^2$)
``$muon\ walls$'' punchthrough detector were placed
behind and at the right side of the calorimeter modules \cite{loc63}.
Basic elements of ``$muon\ walls$'' are plastic scintillator detectors with
dimensions $20 \times 40 \times 2\ cm^3$ which are read-out by 2-inch
photomultipliers $EMI\ 9813KB$.
The tag of given (longitudinal or lateral) leakage is at least one hit in
 corresponding ``$muon\ wall$''.
Due to the number of photoelectrons in any scintillator
counter of walls is roughly
$100$ per minimum ionising particle ``$muon\ walls$''
detected charged particles with high efficiency.
As a result we have for each event 200 values of charges
from $PMT$ properly calibrated \cite{ber} with pedestal subtracted.

\section{Results}

$30$ runs contained $320\ K$
 events with various $Z$ coordinates have been analysed.
The treatment was carried out using program $TILEMON$ \cite{atmon}.

The scintillator detector planes behind and back of the calorimeter
give us  possibility to select the event samples at different conditions:
``no leakage'',  only ``longitudinal leakage'', only ``lateral leakage'',
``longitudinal and lateral leakages'' simultaneously.

In this section the following issues are discussed:
\begin{enumerate}
\item
punchthrough probability,
\item
energy leakage,
\item
the effect of leakage on energy resolution.
\end{enumerate}

First of all we determine the value of punchthrough probability.

\subsection{Longitudinal punchthrough probability}

By definition \cite{loc63}, \cite{fese}, \cite{mer}  
the total hadronic punchthrough
probability is the ratio of the number of events 
giving at least one hit in the
punchthrough detector to the total number of trigger beam particles.
It seems that the information needed is simple: hit or no hit.
But there are some problems in definition of hit (see, for example, 
discussion \cite{fese}).
In Fig.~\ref{fig:f4}
our $ADC$ spectra one of ``$muon\ wall$'' counter
($N^{\b{o}}\ 8$ in Fig.~3 \cite{loc63}) 
in $\mu$ beam (top) and in $\pi$ beam (bottom) are shown.
Spectrum in $\pi$ beam look similar to simulated distribution
for iron-scintillator calorimeter \cite{mer} as obtained 
by Monte Carlo calculations with $GEANT$ (Fig.~14 from \cite{fese}).
The region left from minimum ionising single particle distribution
is related with
the contribution of neutrons as punchthrough particles \cite{fese}.

We used two cuts:
\begin{enumerate}
\item
$ADC_i > ADC_{i}^{L}$,
where   $ADC_{i}^{L}$ --- the beginning of Landau distribution
for $i$-counter,
\item
$ADC_i > 0$
(naturally after pedestal subtraction).
\end{enumerate}
Note that the results of $cut\ 1$ are not so much distinguished from 
a cut used in \cite{loc64} $ADC_i >$ $(<ADC>_i - 3{\sigma}_i)$.

We think that $cut\ 2$ is more correct since it does not reject events with
leakage.
In following for the spectra analysis we use this $cut$.

The results are given in Table~\ref{Tb1}, where in the last raw 
longitudinal punch\-through probability for different cuts corrected 
on value of acceptance of the shower leakage detector 
$(77 \pm 4) \%$ \cite{loc64} are presented.

\begin{table}[tbph]
\caption{
        Percentage of the events and punchthrough probabilities
        for different types of leakages and cuts.
\label{Tb1}}
\begin{center}
\begin{tabular}{|l|c|r|r|c|}
\hline
  Type
& Alias
& $Cut\ 1$
& $Cut\ 2$
& $Cut\ 2/Cut\ 1 - 1$
\\

&  
& \%
& \%
& \%
\\
\hline
\hline
no leak.           & $nl$  & 72.0 & 62.0 & $-14.$\\
\hline
lon.~leak.        & $ll$  & 10.0 & 9.4  &  $-6.$\\
\hline
lat.~leak.         & $lal$ & 14.0 & 22.6 &  61.\\
\hline
lon.~\&~lat.~leak.& $lll$ &  3.6 & 6.0  &  67.\\
\hline
all long.~leak.    & $lol$ & 13.6 & 15.4 &  13. \\
\hline
\hline
punchthrough prob. & $PP$  & 18.$\pm$1. & 20.$\pm$1. & \\
\hline
\end{tabular}
\end{center}
\end{table}

As can be seen from this Table $cut\ 2$ is more soft relative to leakage and
leads to decreasing of the events sample ``no leakage'' and to increasing
of event sample with leakage.
Especially the events sample with ``longitudinal and lateral leakage'' 
are increased (on $67 \%$).

Obtained value of punchthrough probability for $cut\ 1$ 
$(18 \pm 1) \%$
agree with the one from
\cite{loc64}. 
In the case of $cut\ 2$ obtained value $(20 \pm 1) \%$ more correspond to
calculated in \cite{loc64} iron equivalent length $L_{Fe} = 158\ cm$
and the one for a conventional iron-scintillator calorimeter \cite{rd5}.

\subsection{Energy response and leakage}

There are a few methods for evaluating of an energy leakage in calorimetry.
For example, in \cite{aco91} an 
additional ``leakage'' calorimeter was used 
for this purpose special.
In \cite{abram}, \cite{hug} the shower containment was measured 
by using the abundant
longitudinal segmentation information.
Since we do not have such possibilities the following method was used.
We reconstruct the sum of initial energies of showers, $E_{in}$,
by using the detected energies of the event sample ``no leakage'', $E_{nl}$,
and the fraction of these events, $N_{nl}/N_{all}$:
\begin{equation}
\label{eq1}
\sum_{i = 1}^{N_{all}} E_{in}^{i}  = 
\frac{N_{all}}{N_{nl}} \sum_{n = 1}^{N_{nl}} E_{nl}^{n},
\end{equation}
where $N_{all} = N_{nl} + N_{lol}$, 
      $N_{nl}$ --- number of the event sample ``no leakage'',
      $N_{lol}$ --- number of the event sample ``all longitudinal leakage''.

The relative missing leakage energy is equal to:
\begin{equation}
\label{eq2}
L_{r} = 
1 - \frac{\sum_{n = 1}^{N_{nl}} E_{nl}^{n}}{\sum_{i = 1}^{N_{all}}  E_{in}^{i}} - 
\frac{\sum_{l = 1}^{N_{lol}}  E_{ll}^{l}}{\sum_{i = 1}^{N_{all}} E_{in}^{i}} =
\frac{N_{lol}}{N_{all}} \frac{(<E_{nl}> - <E_{ll}>)}{<E_{nl}>}  ,
\end{equation}
where $E_{ll}$ --- energies of the event sample ``longitudinal leakage''.

In Fig.'s~\ref{fig:f5} and \ref{fig:f6}
two-dimensional spectra of energy
responses as a function of $Z$ coordinate and energy $E$ are shown.
Fig.'s~\ref{fig:f7} and \ref{fig:f7a}
show the corresponding energy responses for events with all
$Z$ at different leakage conditions.
To map the energy in $GeV$ scale  the constant equal to 
$100\ GeV$/$<$$E_{nl}$$>$ was used, where $<$$E_{nl}$$> = 514.2\ pC$ 
is the mean energy response for event sample ``no leakage''.
From these figures general behaviour of energy response can be observed.
It is seen that  distributions for event samples ``no leakage'' and
``lateral leakage'' have almost 
$Gaussian$ behaviour, the distribution for event
sample ``longitudinal leakage'' have the clear low
energy tail and in the distribution
for event sample ``lateral leakage'' the  maximum amplitude 
increases with increasing of $Z$.
The obtained mean responses, relative resolutions as well as the values of
leakages and tails are given in Table~\ref{Tb2}, where 
\begin{equation}
\label{eq4}
L = \frac{<E_{nl}> - <E_{i}>}{<E_{nl}>}, 
\end{equation} 
$i =$ {\it ``no leakage'', ``longitudinal leakage'', ``lateral leakage'',
``longitudinal and lateral leakages'', ``all events''}.
The estimate of tail is defined as an excess of the events over 
$Gaussian$ curve
in the region more than one sigma.

\begin{table}[tbph]
\caption{
        Responses, resolutions, leakages and tails 
        for events with different {\it Z}\ in the range from $- 36$ 
        to $20\ cm$.
\label{Tb2}}
\begin{center}
\begin{tabular}{@{}|@{~}l@{~}|@{~}r@{~}|@{~}c@{~}|@{~}c@{~}|@{~}c@{~}|@{~}c@{~}|@{~}c@{~}|@{~}c@{~}|@{}}
\hline
  Type
& \%
& $<${\it E}$>$
& $\frac{\sigma}{<E>}$
& $\frac{\sigma_{i} - \sigma_{nl}}{\sigma_{nl}}$
& $L$
%& $\frac{<E_{nl}> - <E_{i}>}{<E_{nl}>}$
& Low tail
& High tail
\\

& Events
& $GeV$
& \%
& \%
& \%
& \%
& \%
\\
\hline
\hline
no leak.            & 62.0 & 100. & 7.4  & 0.0  & 0.0 & 0.0  & 2.6$\pm$0.05\\
\hline
lon.~leak.         & 9.4  & 91.0 & 10.4 & 41.  & 9.0 & 7.1$\pm$0.2 & 0.0 \\
\hline
lat.~leak.          & 22.6 & 96.8 & 7.2  & $-1.9$ & 3.2 & 0.0 & 1.1$\pm$0.05 \\
\hline
lon.\ \& lat.leak.\ &  6.0 & 88.3 & 10.7 & 45.  & 11.7 & 6.1$\pm$0.2 & 0.0\\
\hline
all events          & 100. & 97.7 & 8.3  & 13.  & 2.3 & 1.5$\pm$0.03 & 1.3$\pm$0.02 \\
\hline
\end{tabular}
\end{center}
\end{table}

\begin{table}[tbph]
\caption{
        Responses, resolutions, leakages and tails for the events with
        $Z = -8\ cm$ at various leakage conditions.
\label{Tb3}}
\begin{center}
\begin{tabular}{@{}|@{~}l@{~}|@{~}r@{~}|@{~}c@{~}|@{~}c@{~}|@{~}c@{~}|@{~}c@{~}|@{~}c@{~}|@{~}c@{~}|@{}}
\hline
  Type
& \%
& $<${\it E}$>$
& $\frac{\sigma}{<E>}$
& $\frac{\sigma_{i} - \sigma_{nl}}{\sigma_{nl}}$
& $L$
%& $\frac{<E_{nl}> - <E_{i}>}{<E_{nl}>}$
& Low tail
& High tail
\\

& Events
& $GeV$
& \%
& \%
& \%
& \%
& \%
\\
\hline
\hline
no leak.            & 71.3 & 100. & 7.3  & 0.0   & 0.0   & 0.0 & 2.7$\pm$0.2\\
\hline
lon.~leak.         & 11.1 & 91.0 & 9.9  & 35.   & 9.0   & 6.7$\pm$0.7 & 0.0 \\
\hline
lat.~leak.          & 14.2 & 98.8 & 7.1  & $-4.$ & 1.2   & 0.0  & 1.7$\pm$0.3 \\
\hline
lon.\ \& lat.leak.\ & 3.3  & 89.5 & 8.8  & 20.   & 10.5   & 12.$\pm$2. & 0.0 \\
\hline
all events          &100   & 98.4 & 9.8  & 7.9   & 1.6   & 1.5$\pm$0.1 & 1.3$\pm$0.1 \\
\hline
\end{tabular}
\end{center}
\end{table}

Fig.'s~\ref{fig:f8} and \ref{fig:f8a} show the energy distributions for 
event samples with various leakage conditions at $Z= - 8\ cm$.
The characteristics of these distributions are given in Table~\ref{Tb3}.
The event samples --- ``any leakage'', ``longitudinal leakage'', 
``longitudinal and lateral leakages'' have the low energy tails,
the event
samples --- ``no leakage'' and ``lateral leakage'' have the high 
energy tails.
The events sample with leakage naturally have the low energy tail.
The high energy tail in the event sample
``no leakage'' was explained in \cite{amaldi} by contribution of showers with unusually large
electromagnetic component. 
The unexpected high energy tail in the event sample ``lateral leakage'' may be explained 
as these events are the events
of type ``no leakage'' with some leakage unsufficient to cut the high energy tail.

In Fig.~\ref{fig:f9} are shown the 
mean energy responses for events with different types of leakage obtained by 
averaging of energy spectra (top) and $Gaussian$ fits (bottom)
as a function of $Z$ coordinate at different leakage conditions.
In Table~\ref{Tb4}  are given the results of averaging of these dependences in 
the uniformity ranges.

\begin{table}[tbph]
\caption{
        Responses and  resolutions for the events 
        at various leakage conditions.
\label{Tb4}}
\begin{center}
\begin{tabular}{@{}|@{~}l@{~}|@{~}c@{~}|@{~}c@{~}|@{~}c@{~}|@{~}c@{~}|@{~}c@{~}|@{~}c@{~}|@{}}
\hline
  Type
& $<${\it E}$>$
& $E_{G}$
& $\frac{RMS}{<E>}$
& $\frac{\sigma}{E_{G}}$
& $L$
%& $\frac{<E_{nl}> - <E_{i}>}{<E_{nl}>}$
& $\frac{\sigma_{i} - \sigma_{nl}}{\sigma_{nl}}$
\\
  
& $GeV$
& $GeV$
& \%
& \%
& \%
& \%
\\
\hline
\hline
no leak.      & 100.$\pm$0.02 & 99.7$\pm$0.02 & 8.0$\pm$0.02  &  7.4$\pm$0.02 
& 0.0 & 0.0
\\
\hline
lon.~lk.    & 91.1$\pm$0.12 & 94.0$\pm$0.08 & 16.2$\pm$0.1 &  9.7$\pm$0.07 
& 8.9$\pm$0.1 & 31.0$\pm$0.8 
\\
\hline
lat.~lk.$^*$& 98.3$\pm$0.06 & 98.1$\pm$0.06 & 7.6$\pm$0.03  &  7.3$\pm$0.03 
& 1.7$\pm$0.1 & -$1.4\pm0.5$ 
\\
\hline
all ev.$^*$& 98.5$\pm$0.02 & 98.8$\pm$0.02 & 9.9$\pm$0.02  &  8.0$\pm$0.02 
& 1.5$\pm$0.1 & 8.1$\pm$0.4 
\\
\hline
\multicolumn{7}{l}{$^*$ For events with $Z < 5\ cm$.}
\end{tabular}
\end{center}
\end{table}

The fraction of the energy
leaking out from the backward side of this 
calori\-me\-ter calculated by the 
formula (\ref{eq2})  amounts to $(1.8\pm0.03) \%$ 
and agrees with the value $1.73 \%$ for 
$L_{Fe} = 158\ cm$ measured in \cite{hug}.

It should be noted that $15 \%$ of the events have the $9 \%$ energy longitudinal
leakage and  $1 \%$ of the events $50 \%$ of energy 
($\approx 50\ GeV$) leaking out at average.
The latter estimate is extracted from the low energy tail in Fig.~\ref{fig:f7a} (top).
This fact must be taken into account in searching of new particles in
future $LHC$ experiments.

We also considered the question concerning nonuniformity 
response of calori\-meter.
As can be seen in Fig.~\ref{fig:f9} the energy response as a function of 
$Z$ coordinate from event sample ``no leakage'' is more uniform than the one
for other event types.
It is allows to estimate the more extended range of uniformity
(from $- 36\ cm$ to $20\ cm$) than in \cite{ber}
which appears equal to $0.9 \%$ ($RMS$).

\subsection{Influence of leakage on the energy resolution}

Fig.~\ref{fig:f10}
shows the relative energy resolutions 
obtained by $Gaussian$ fitting of spectra
(top) and the relative energy resolutions ($RMS$/$<${\it E}$>$)
obtained by averaging of spectra (bottom)
as a function of $Z$ coordinate at different leakage conditions.
Fig.~\ref{fig:f11} shows the same normalised to average value of 
($\sigma /E_{G}$) over the  uniformity range for events without leakage.
One can see that due to the tails the resolutions obtained by
averaging are much greater (approximately in two times for
events with  longitudinal leakage) than ones obtained by 
$Gaussian$ fitting.
The results of averaged by $Z$ in their uniformity range 
of Fig.'s~\ref{fig:f9}, \ref{fig:f10}, \ref{fig:f11}
are given in Table~\ref{Tb4}.
As can be seen 
longitudinal energy leakage amounts  $9 \%$, but deterioration of 
energy resolution for the same case  $\sigma /E_{G}$ amounts $31 \%$.
The general degradation of the resolution 
with increasing of leakage is in agreement with earlier observations
\cite{aco91}, \cite{amaldi}, \cite{barr}, \cite{hen}.
Moreover, our energy resolution degradation 
$\frac{(\sigma_{ll} - \sigma_{nl})}{\sigma_{ll}} = 24 \%$ 
is in reasonable agreement
with the parameterisation proposed by \cite{fnal} on the basis of the data
from $CITF$ collaboration \cite{scul}:
\begin{equation}
\label{eq3}
\frac{(\sigma_{l} - \sigma_{0})}{\sigma_{l}} = 
0.9 \cdot \sqrt{\frac{<E_{0}> - <E_{l}>}{<E_{0}>}},
\end{equation}
where  
$<E_{0}> = <E_{nl}>$ and $\sigma_{0} = \sigma_{nl}$ --- 
energy and energy resolution for events without leakage, 
$<E_{l}> = <E_{ll}>$ and $\sigma_{l} = \sigma_{ll}$ --- 
energy and energy resolution for events with ``longitudinal leakage''.
In our case for the value of energy resolution degradation 
from  (\ref{eq3}) we obtain  $27 \%$.

In the case of lateral leakage the unexpected inverse behaviour is observed:
energy leakage leads to some improving of the resolution.
Let us  consider this in more detail. 
In Fig.~\ref{fig:f12} two distributions of the lateral leakage are
shown: lateral energy leakage (top), energy resolution ($\sigma /E_{G}$) 
(bottom)
for the event sample with lateral leakage as a function of $Z$ coordinate.

Fig.~\ref{fig:f13} presents the energy resolution as a function of lateral
leakage for this event sample.
As can be seen the energy resolution improves with increasing lateral
energy leakage at least to the value of  lateral 
energy leakage equal to $6 \%$ at $Z = 18\ cm$ where  energy resolution is
improving to $18 \%$.

This phenomenon  can be explained as follows. 
The hadronic shower
consists of electromagnetic and pure hadronic parts and the
electromagnetic part in lateral direction places 
in the central core  \cite{barr}, \cite{aco92}.
So by cutting some lateral hadronic part we ``improve'' the shower properties,
make it less fluctuating.
However this may be the specific property of our calorimeter.

\section{Conclusions}

We have investigated the
hadronic shower longitudinal and lateral leakages and its effect
on the pion response and energy resolution on the basis of $100\ GeV$
pion beam data at incidence angle
$\Theta = 10^{o}$ at impact points $Z$ in the range from $- 36$ to $20\ cm$.

\bigskip

Some results are following:
\begin{itemize}
\item
The fraction of the energy of $100\ GeV$ pions at $\Theta = 10^{o}$
leaking out at the back of this calorimeter amounts to $1.8 \%$ 
and agrees with the one for a conventional 
iron-scintillator calorimeter. 
\item
Unexpected behaviour of the energy resolution as a function of leakage is
observed: $6 \%$ lateral leakage leads to $18 \%$
improving of energy resolution in compare to events with the 
showers without leakage.
\item
The measured value of longitudinal punchthrough probability $(20 \pm 1) \%$
agrees with the one for a conventional iron-scintillator calorimeter with the 
same nuclear interaction length thickness and with the earlier measurement 
\cite{loc64}.
It also more correspond 
to calculated in \cite{loc64} iron equivalent length $L_{Fe} = 158\ cm$.
\end{itemize}

\section{Acknowledgements}

This work is the result of the efforts of many people from 
{\it ATLAS Collaboration}.
The authors are greatly indebted to all Collaboration
for their test beam setup and data taking.

Authors are grateful {\it Peter Jenni} and 
{\it Nikolai Russakovich} for their attention and support of this work.
We are indebted to {\it Martin Bosman}, {\it Mateo Ca\-va\-l\-li-Sforza},
{\it Ana Henriques} and {\it Stanislav Tokar} for the valuable discussions 
and constructive advices.
The authors would like to thank {\it Stanislav Ne\-me\-cek}  for giving the numerical 
value of acceptance for $backward$ ``{\it muon wall}'' for $\pi$ at $100\ GeV$
and valuable discussions.

%\newpage

}

%\newpage

\begin{figure*}[tbph]
     \begin{center}
        \begin{tabular}{c}
        %\hline
        \mbox{\epsfig{figure=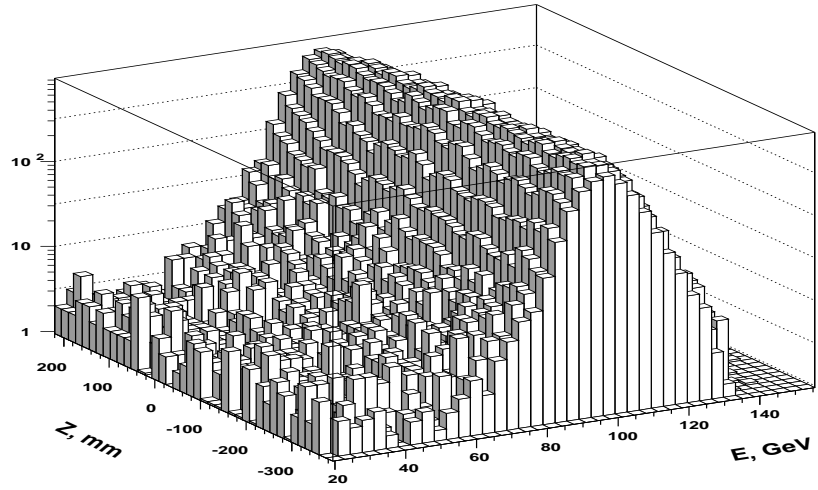,width=0.95\textwidth,height=0.4\textheight}} \\
        %\hline
        %\hline
        \mbox{\epsfig{figure=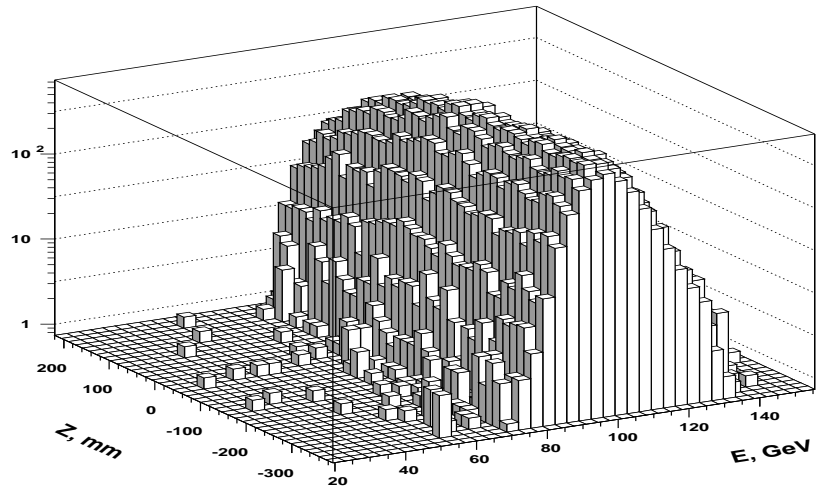,width=0.95\textwidth,height=0.4\textheight}} \\
        %\hline
        \end{tabular}
     \end{center}
      \caption{
        Two dimensional spectrum of energy response as a function of $Z$
        coordinate and energy $E$ for various leakage conditions:
        all events (top),
        no leakage (bottom).
       \label{fig:f5}}
\end{figure*}

\begin{figure*}[tbph]
     \begin{center}
        \begin{tabular}{c}
        %\hline
        \mbox{\epsfig{figure=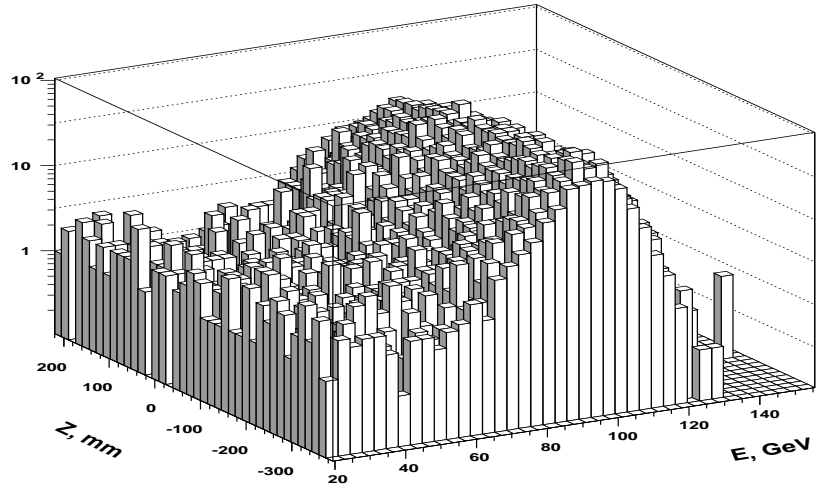,width=0.95\textwidth,height=0.4\textheight}} \\
        %\hline
        %\hline
        \mbox{\epsfig{figure=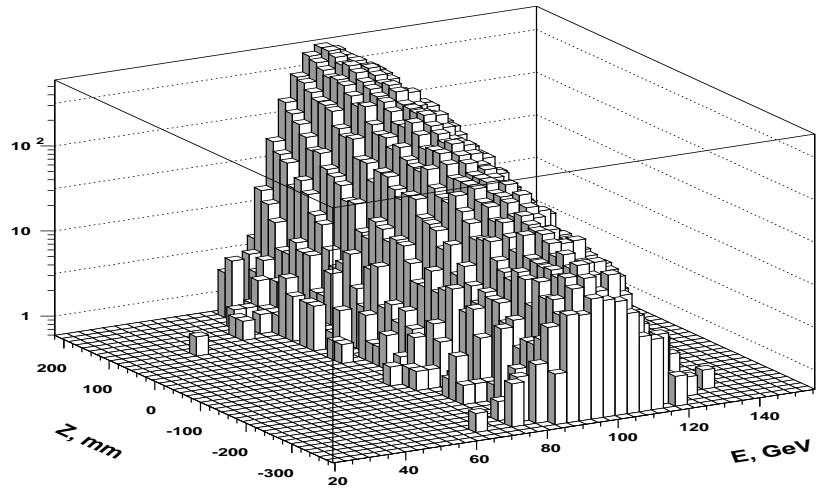,width=0.95\textwidth,height=0.4\textheight}} \\
        %\hline
        \end{tabular}
     \end{center}
      \caption{
        Two dimensional spectrum of energy response as a function of $Z$
        coordinate and energy $E$ for various leakage conditions:
        longitudinal leakage (top),
        lateral leakage (bottom).
       \label{fig:f6}}
\end{figure*}

\begin{figure*}[tbph]
     \begin{center}
        \begin{tabular}{c}
        %\hline
        \mbox{\epsfig{figure=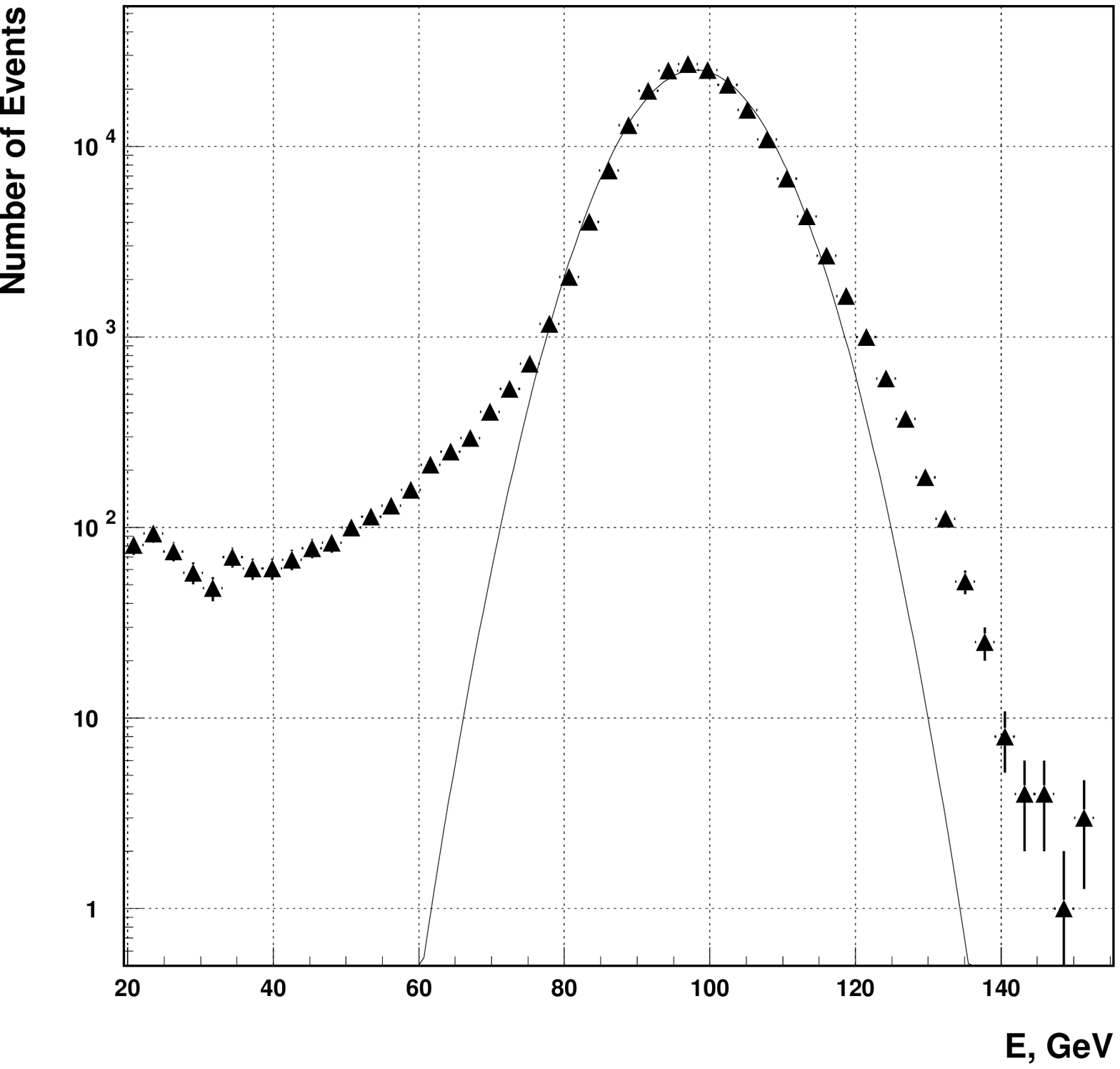,width=0.95\textwidth,height=0.4\textheight}} \\
        %\hline
        %\hline
        \mbox{\epsfig{figure=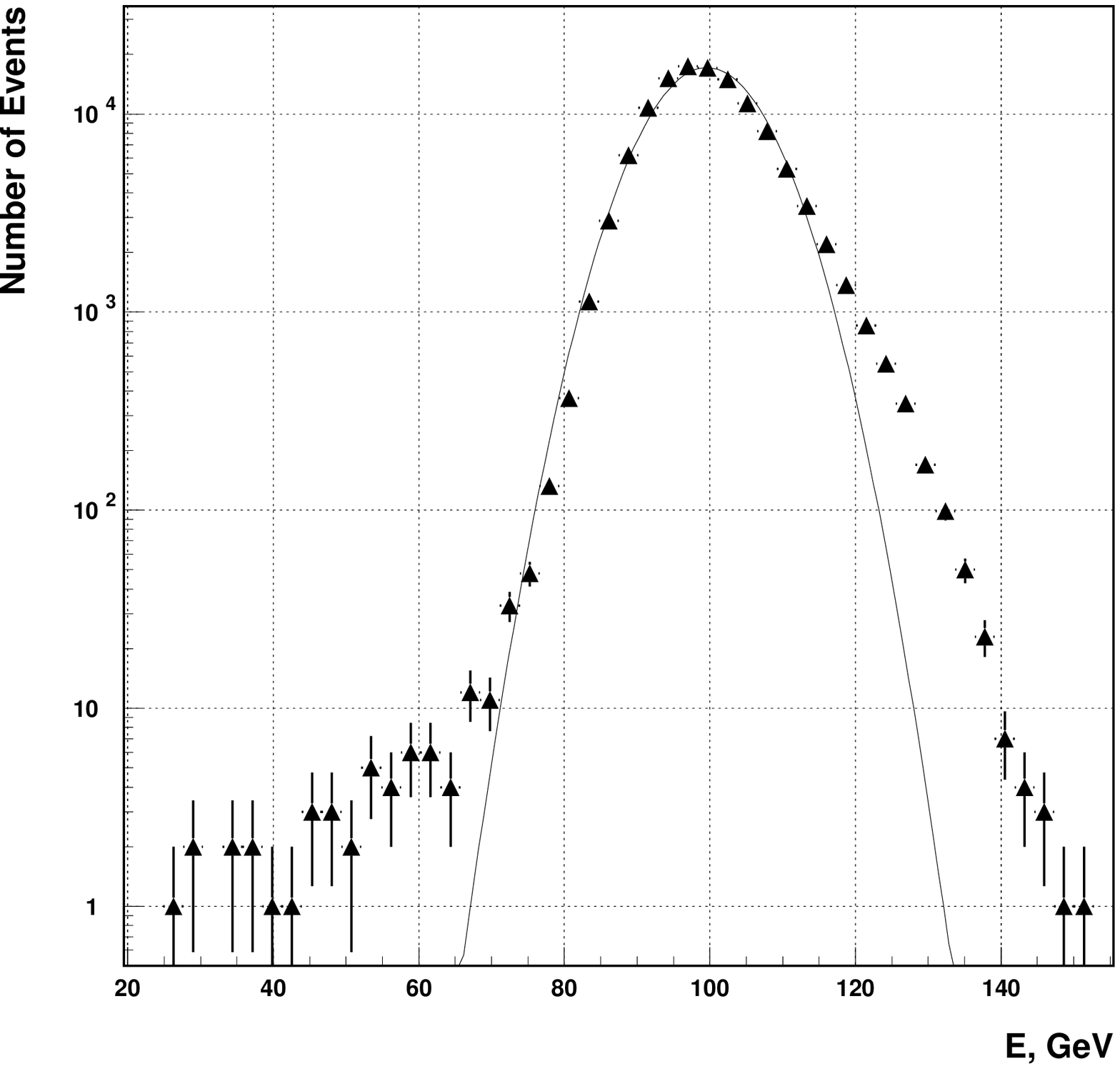,width=0.95\textwidth,height=0.4\textheight}} \\
        %\hline
        \end{tabular}
     \end{center}
      \caption{
        Energy responses for all $Z$ at a different leakage conditions:
        all events (top),
        no leakage (bottom).
       \label{fig:f7}}
\end{figure*}

\begin{figure*}[tbph]
     \begin{center}
        \begin{tabular}{c}
        %\hline
        \mbox{\epsfig{figure=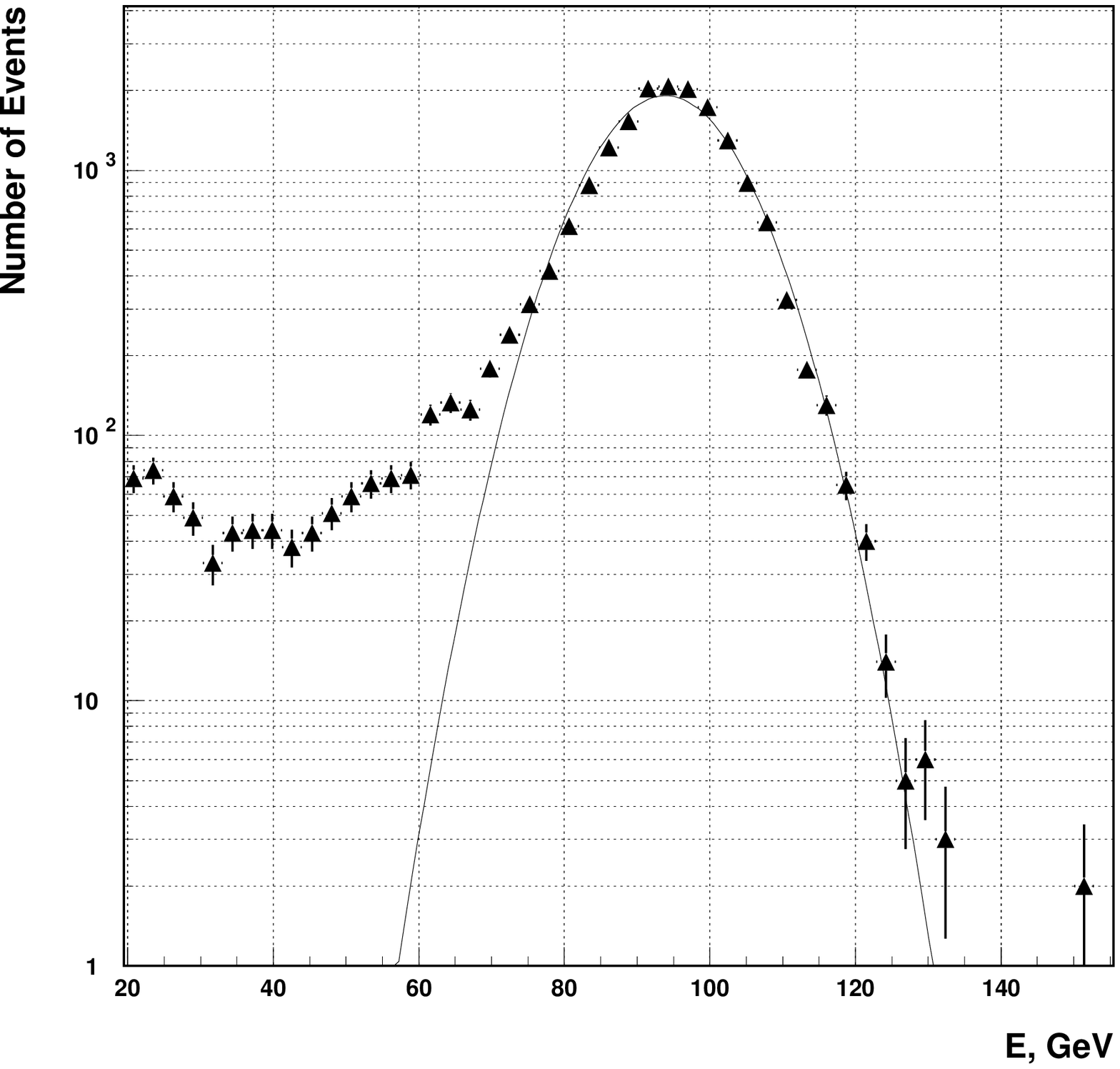,width=0.95\textwidth,height=0.4\textheight}} \\
        %\hline
        %\hline
        \mbox{\epsfig{figure=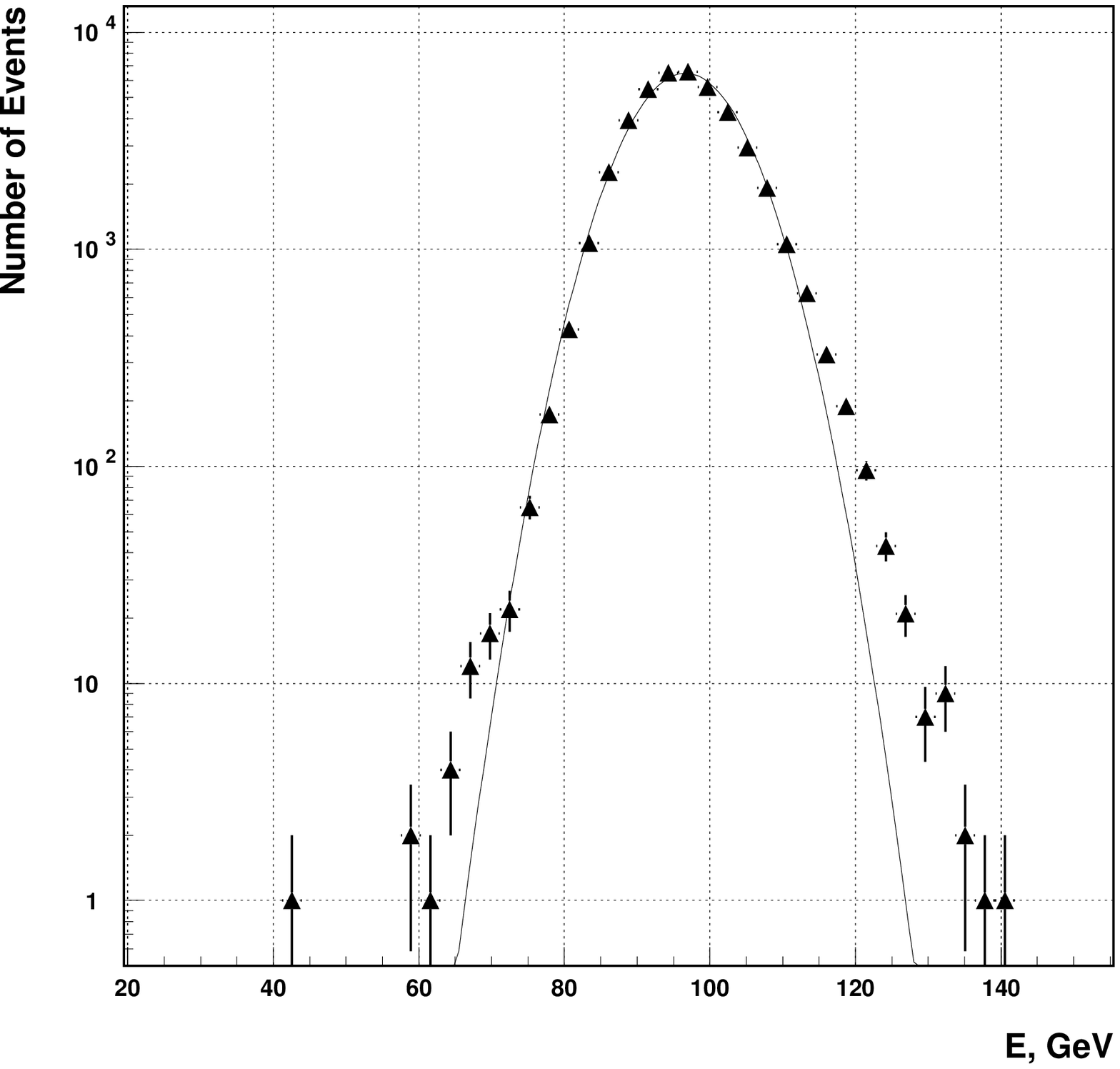,width=0.95\textwidth,height=0.4\textheight}} \\
        %\hline
        \end{tabular}
     \end{center}
      \caption{
        Energy responses for $Z$ at a different leakage conditions:
        longitudinal leakage (top),
        lateral leakage (bottom).
       \label{fig:f7a}}
\end{figure*}

\begin{figure*}[tbph]
     \begin{center}
        \begin{tabular}{c}
        %\hline
        \mbox{\epsfig{figure=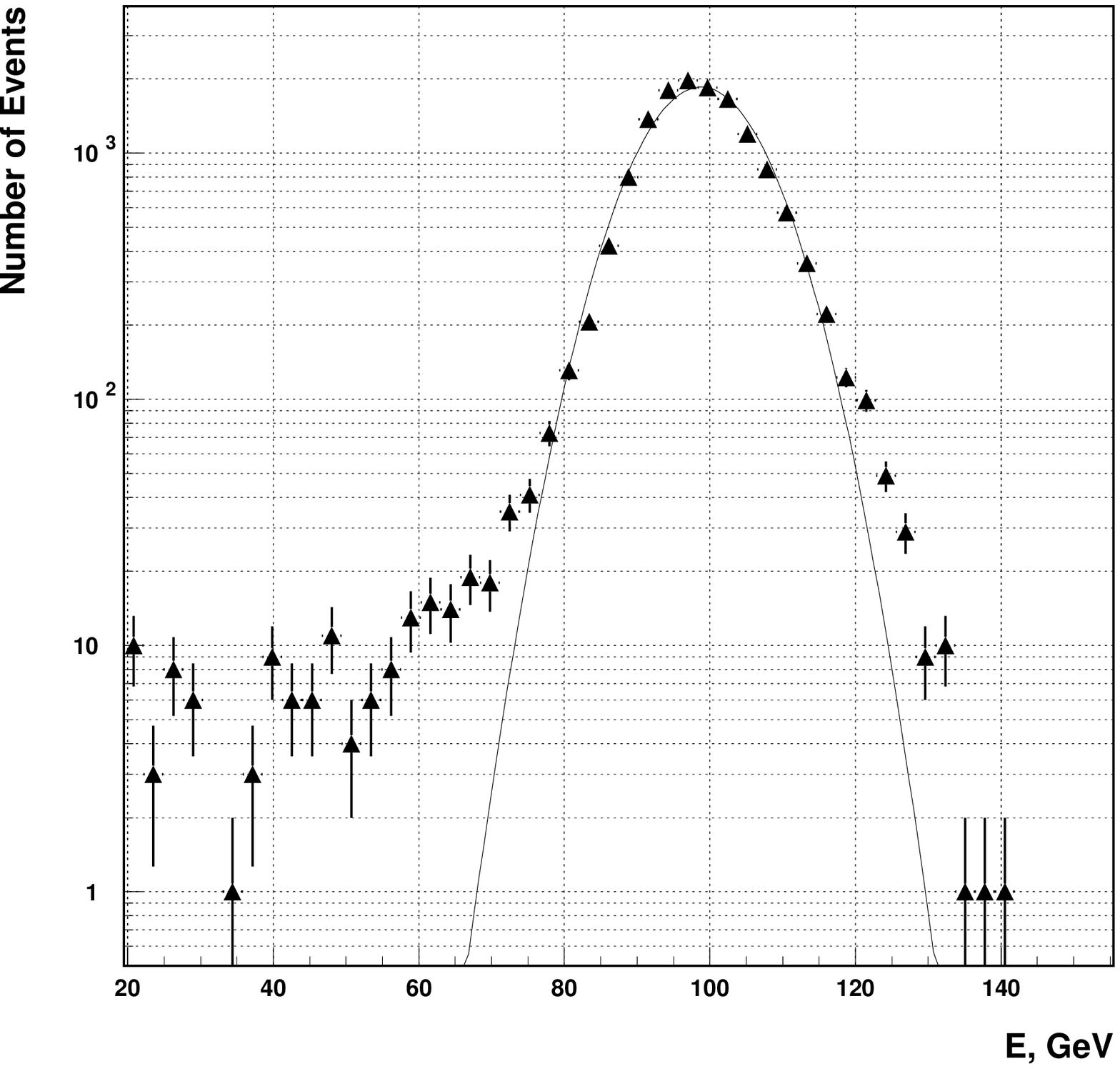,width=0.95\textwidth,height=0.4\textheight}} \\
        %\hline
        %\hline
        \mbox{\epsfig{figure=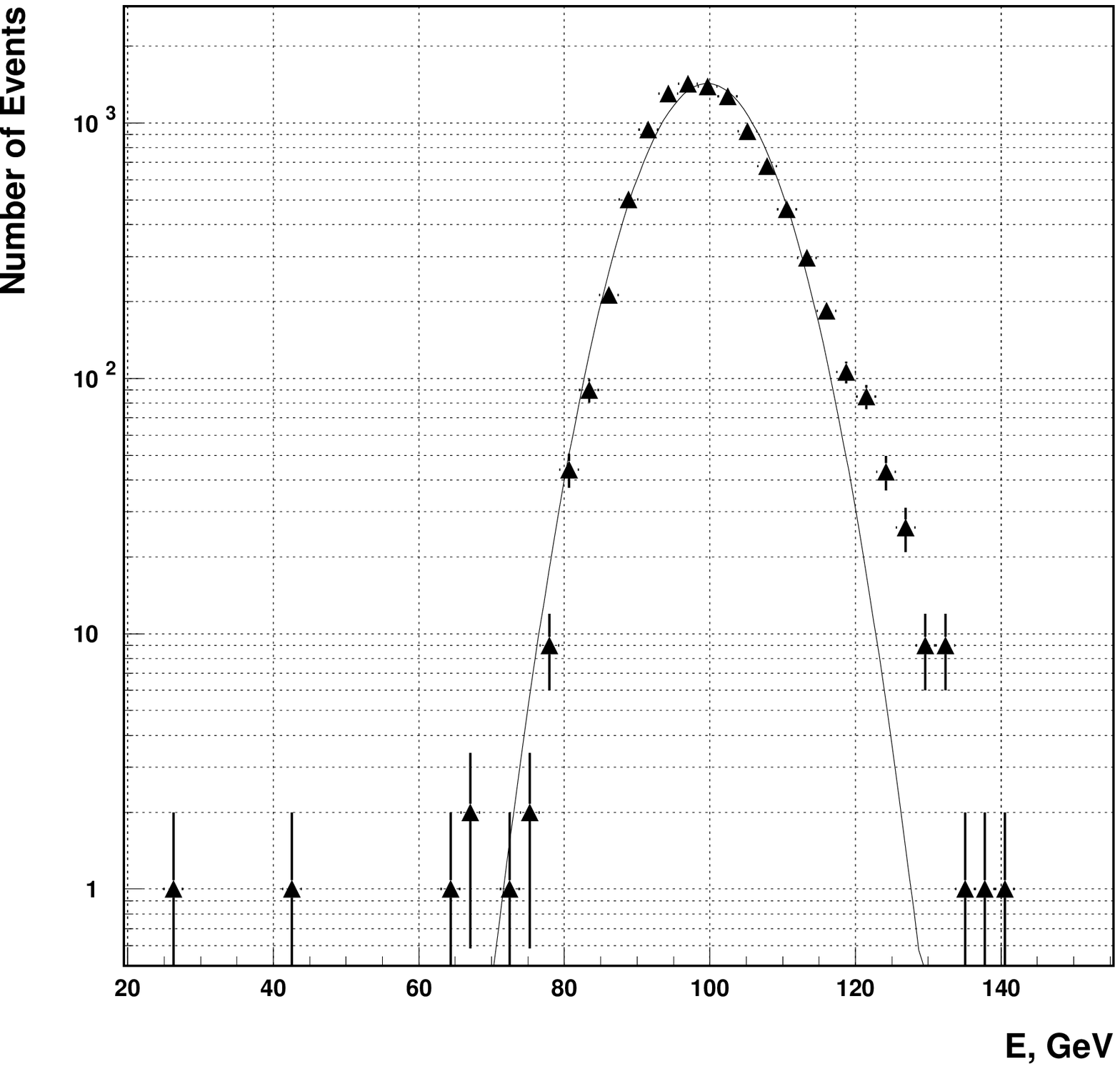,width=0.95\textwidth,height=0.4\textheight}} \\
        %\hline
        \end{tabular}
     \end{center}
      \caption{
        Energy responses for all $Z = - 8\ cm$ at 
        a different leakage conditions:
        all events (top),
        no leakage (bottom).
       \label{fig:f8}}
\end{figure*}

\begin{figure*}[tbph]
     \begin{center}
        \begin{tabular}{c}
        %\hline
        \mbox{\epsfig{figure=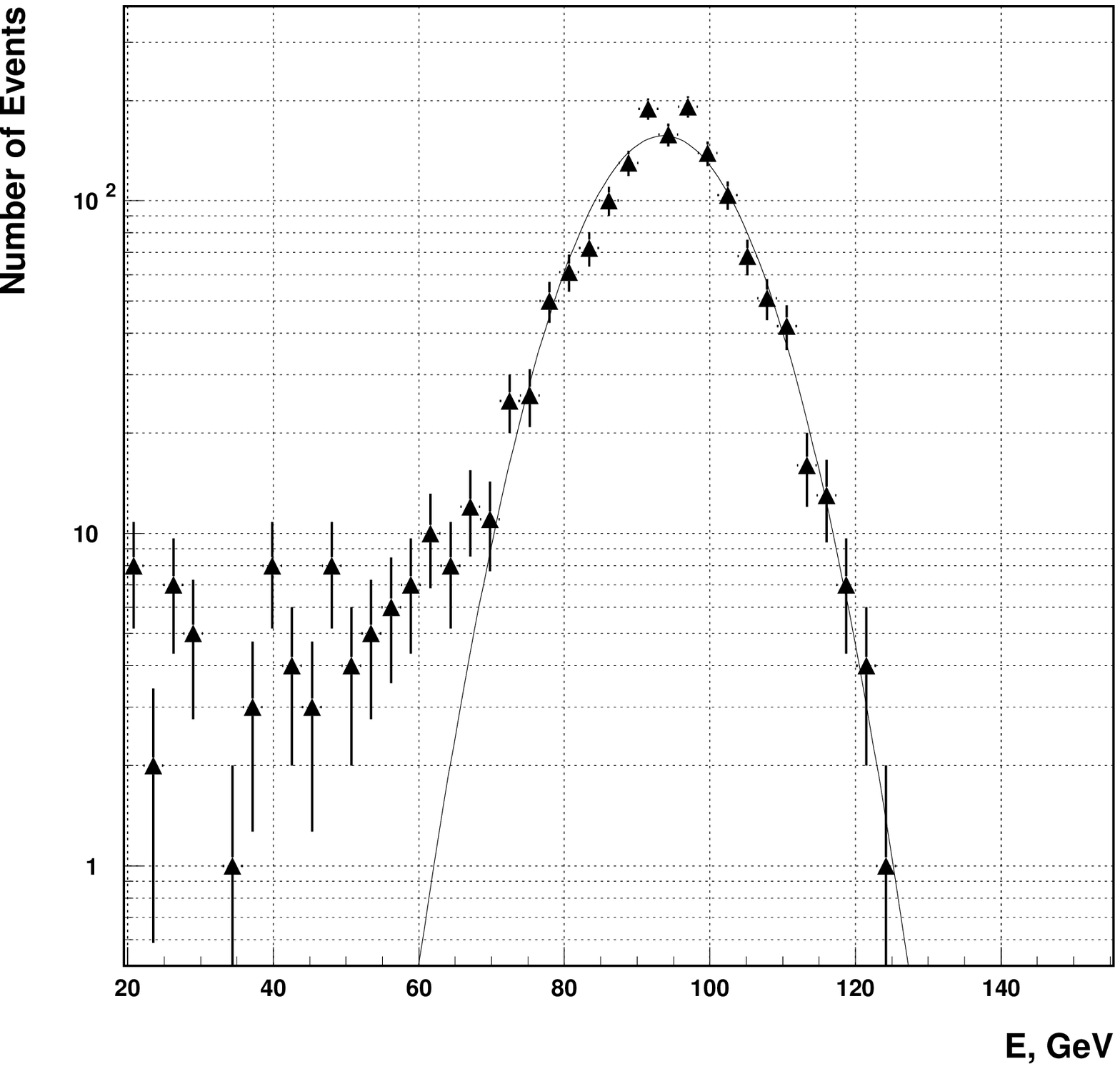,width=0.95\textwidth,height=0.4\textheight}} \\
        %\hline
        %\hline
        \mbox{\epsfig{figure=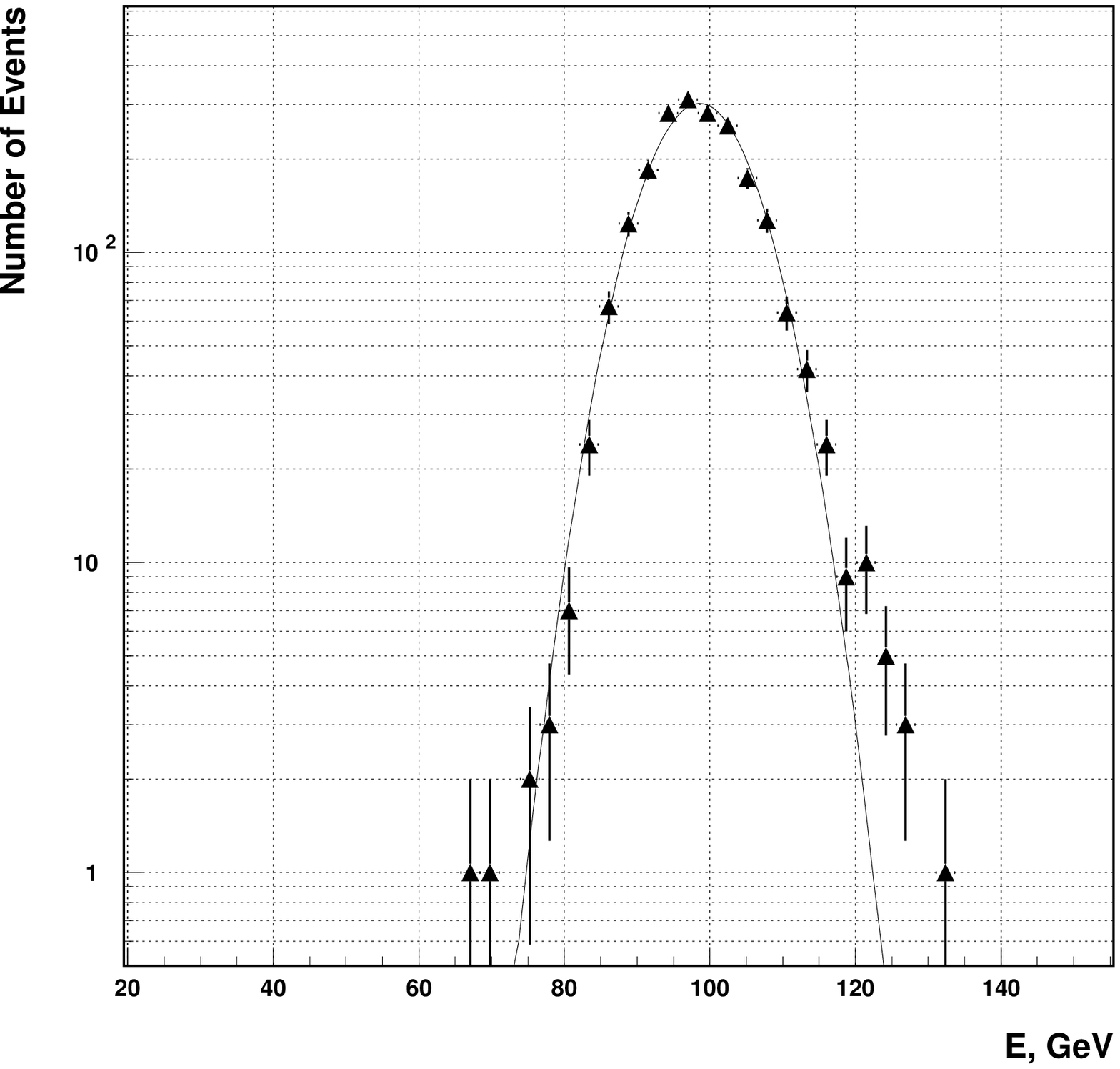,width=0.95\textwidth,height=0.4\textheight}} \\
        %\hline
        \end{tabular}
     \end{center}
      \caption{
        Energy responses for all $Z = - 8\ cm$ at 
        a different leakage conditions:
        longitudinal leakage (top),
        lateral leakage (bottom).
       \label{fig:f8a}}
\end{figure*}

\begin{figure*}[tbph]
     \begin{center}
        \begin{tabular}{c}
        %\hline
        \mbox{\epsfig{figure=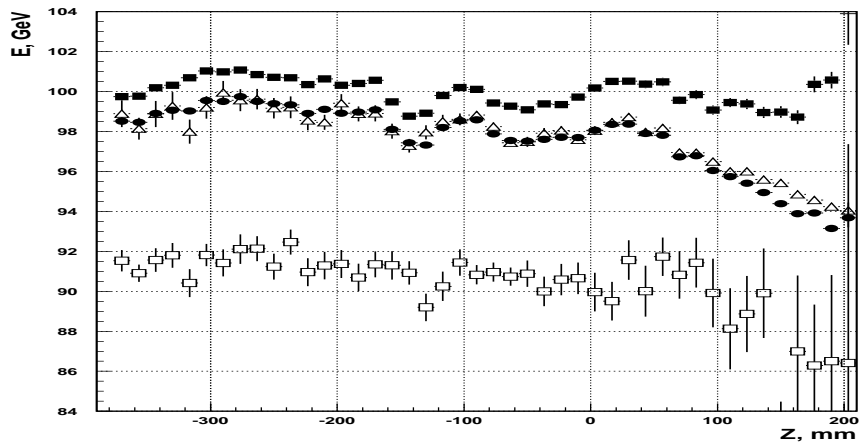,width=0.95\textwidth,height=0.35\textheight}} \\
        %\hline
        %\hline
        \mbox{\epsfig{figure=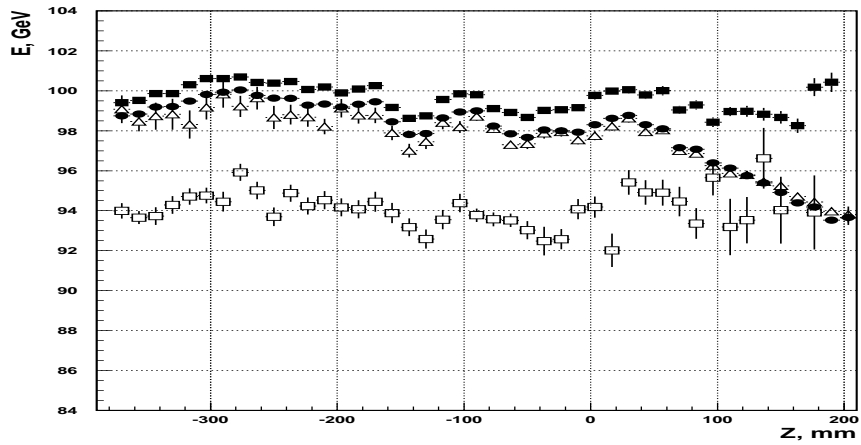,width=0.95\textwidth,height=0.35\textheight}} \\
        %\hline
        \end{tabular}
     \end{center}
      \caption{
        Mean energy responses obtained by averaging of spectrum (top) 
        and $Gaussian$ fitting (bottom)
        as a function of $Z$ coordinate at different leakage conditions:
        a) black square --- no leakage,
        b) open square --- longitudinal leakage,
        c) open triangle --- lateral leakage,
        d) black circle --- all events.
       \label{fig:f9}}
\end{figure*}

\begin{figure*}[tbph]
     \begin{center}
        \begin{tabular}{c}
        %\hline
        \mbox{\epsfig{figure=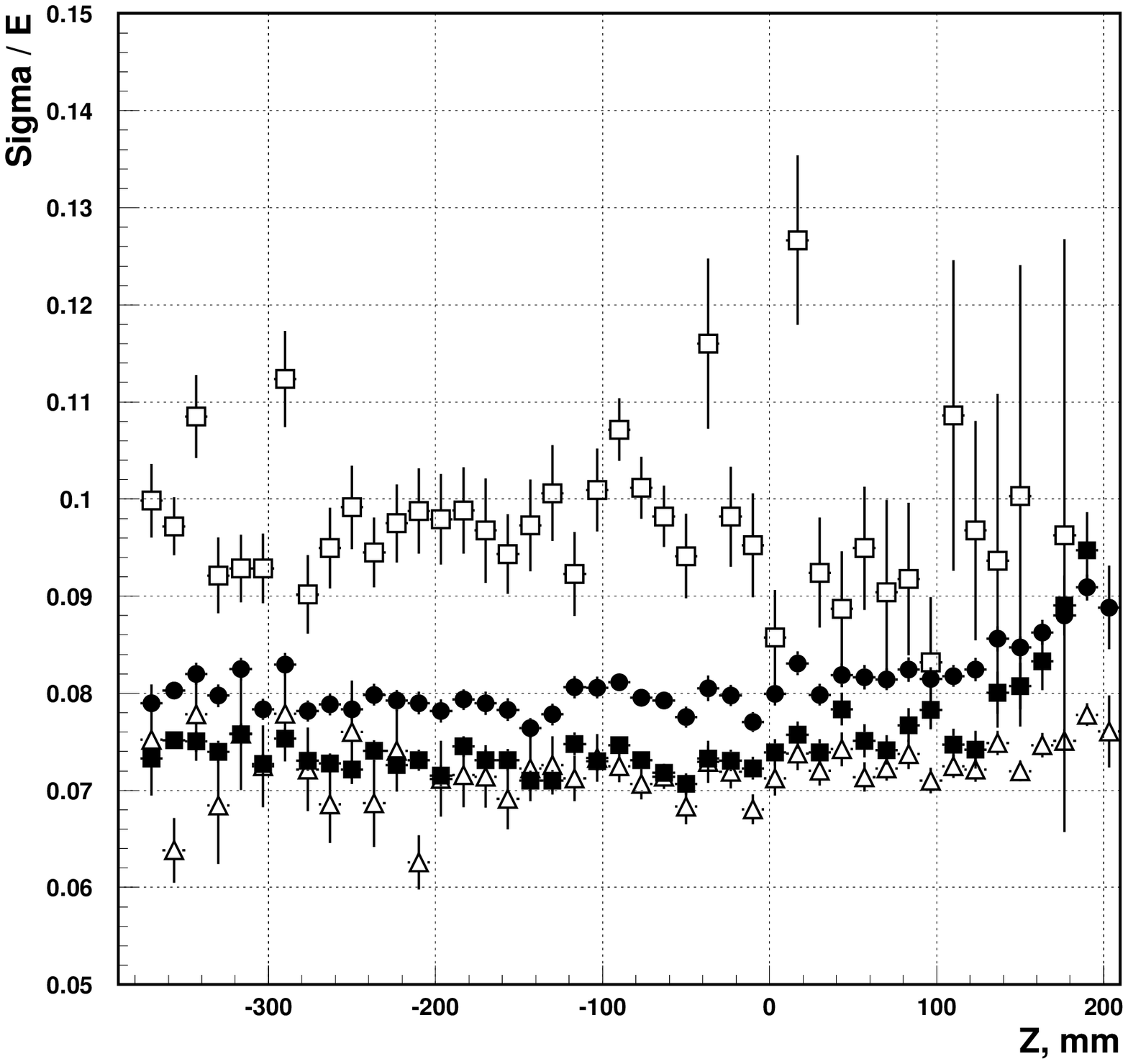,width=0.95\textwidth,height=0.35\textheight}} \\
        %\hline
        %\hline
        \mbox{\epsfig{figure=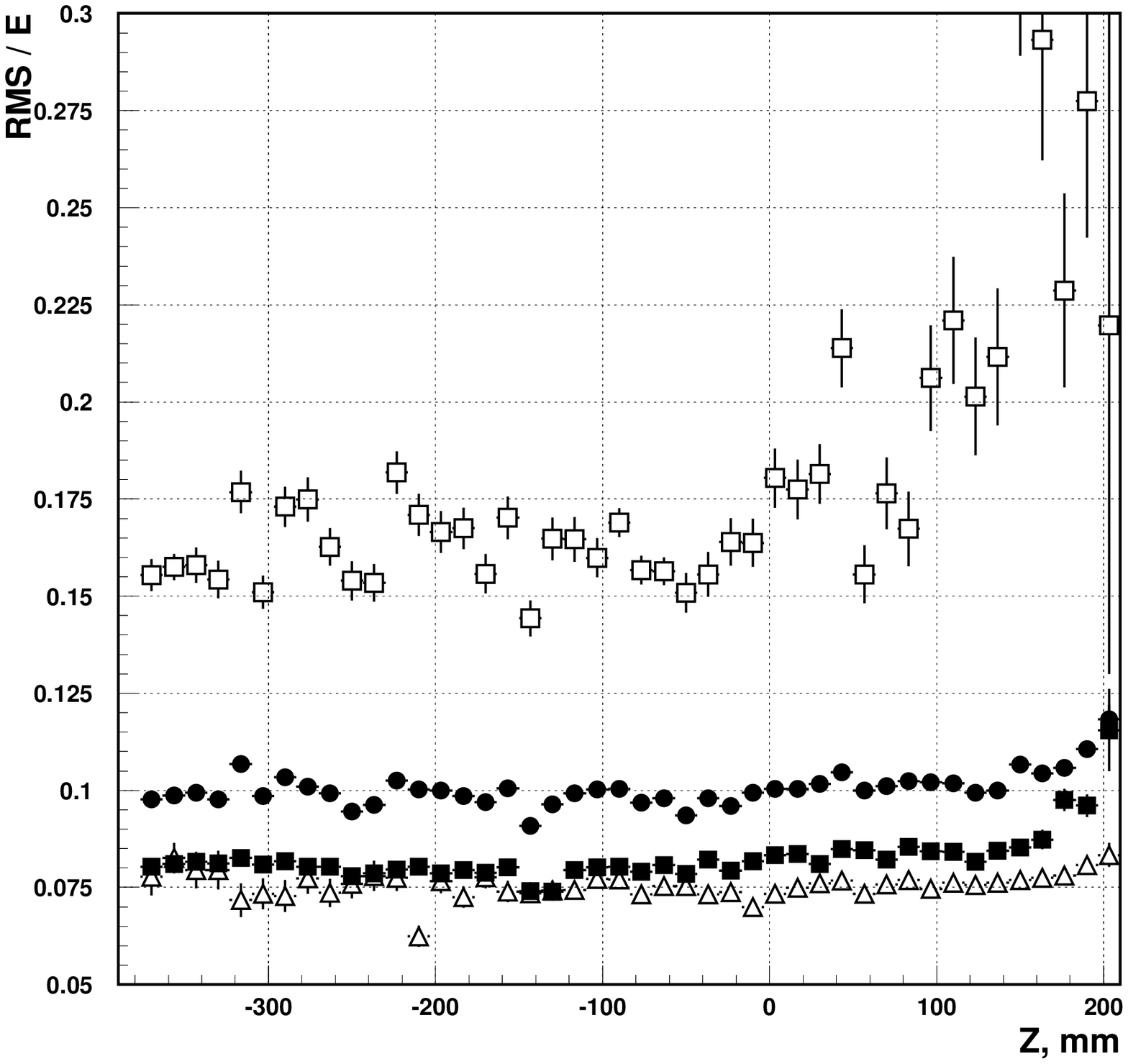,width=0.95\textwidth,height=0.35\textheight}} \\
        %\hline
        \end{tabular}
     \end{center}
      \caption{
        Energy resolutions ($\sigma /E_{G}$) obtained by $Gaussian$
        fitting (top)
        and energy resolutions ($RMS/<E>$) obtained 
        by averaging of spectrum 
        (bottom)
        as a function of $Z$ coordinate at different leakage conditions:
        a) black square --- no leakage,
        b) open square --- longitudinal leakage,
        c) open triangle --- lateral leakage,
        d) black circle --- all events.
       \label{fig:f10}}
\end{figure*}

\begin{figure*}[tbph]
     \begin{center}
        \begin{tabular}{c}
        %\hline
        \mbox{\epsfig{figure=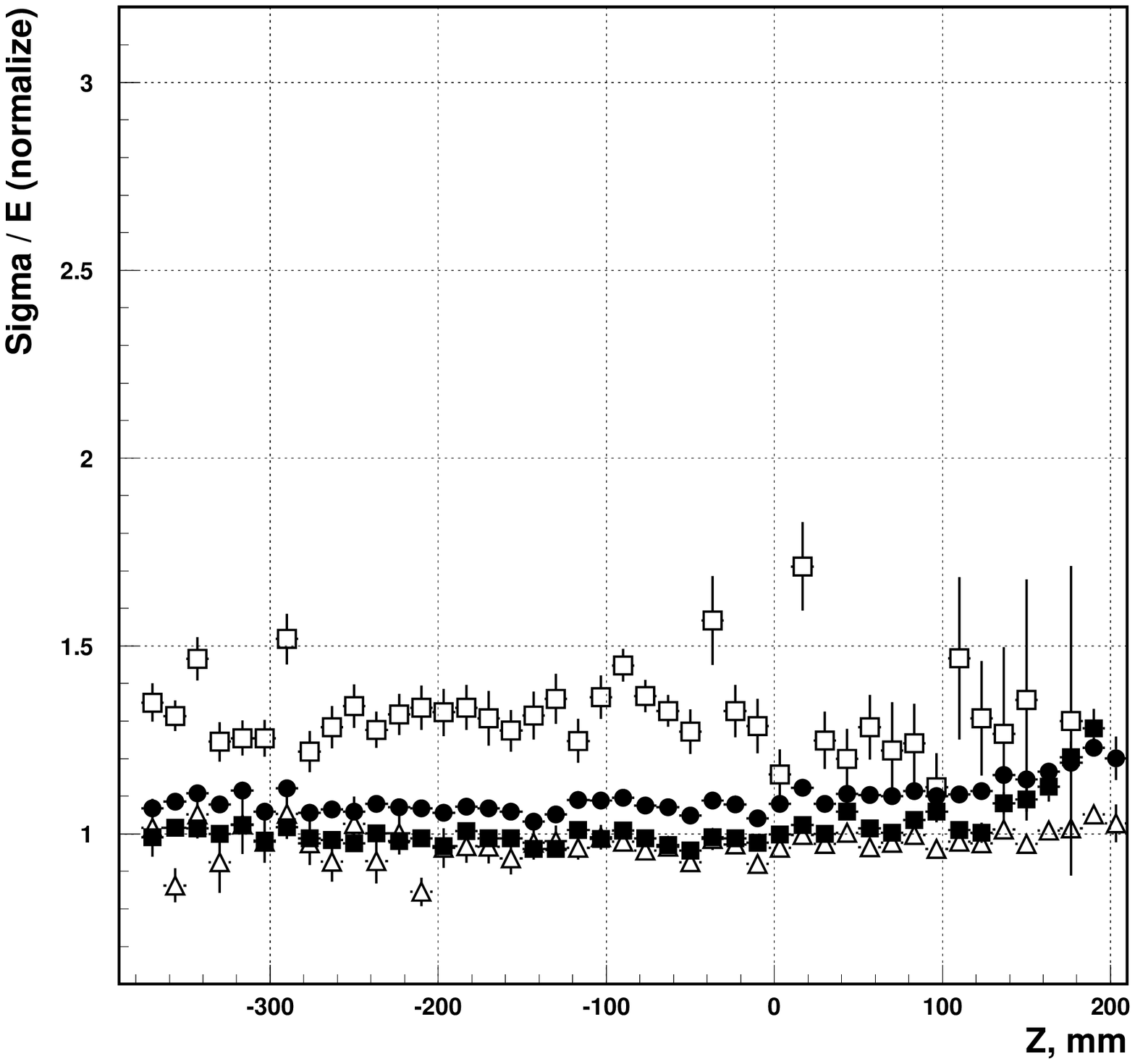,width=0.95\textwidth,height=0.35\textheight}} \\
        %\hline
        %\hline
        \mbox{\epsfig{figure=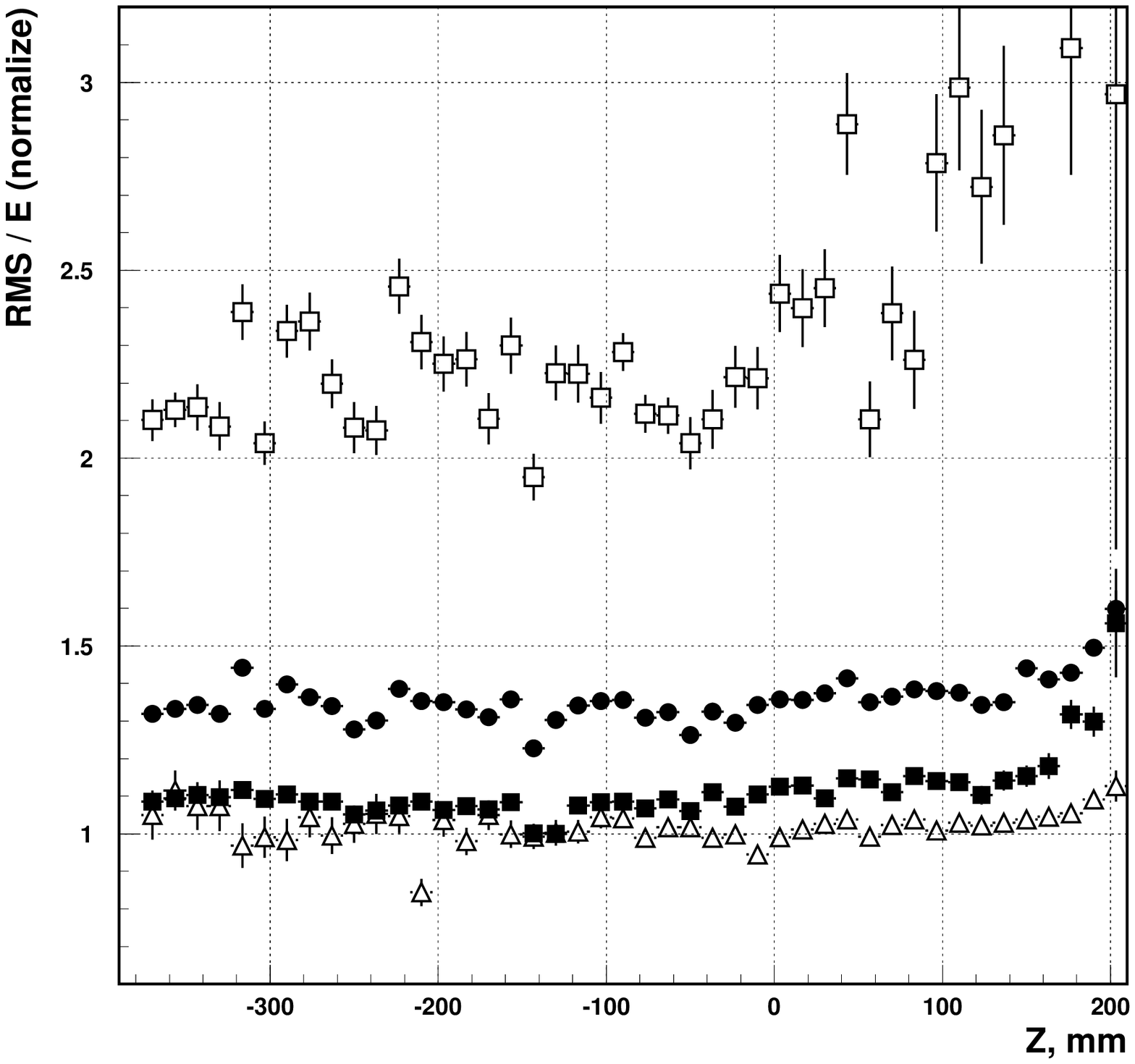,width=0.95\textwidth,height=0.35\textheight}} \\
        %\hline
        \end{tabular}
     \end{center}
      \caption{
        Normalised energy resolutions ($\sigma /E_{G}$) obtained by 
        Gaussian fitting (top)
        and normalised energy resolutions ($RMS /$ $<E>$) 
        obtained by averaging 
        of spectrum (bottom)
        as a function of $Z$ coordinate at different leakage conditions:
        a) black square --- no leakage,
        b) open square --- longitudinal leakage,
        c) open triangle --- lateral leakage,
        d) black circle --- all events.
       \label{fig:f11}}
\end{figure*}

\begin{figure*}[tbph]
     \begin{center}
        \begin{tabular}{c}
        %\hline
        \mbox{\epsfig{figure=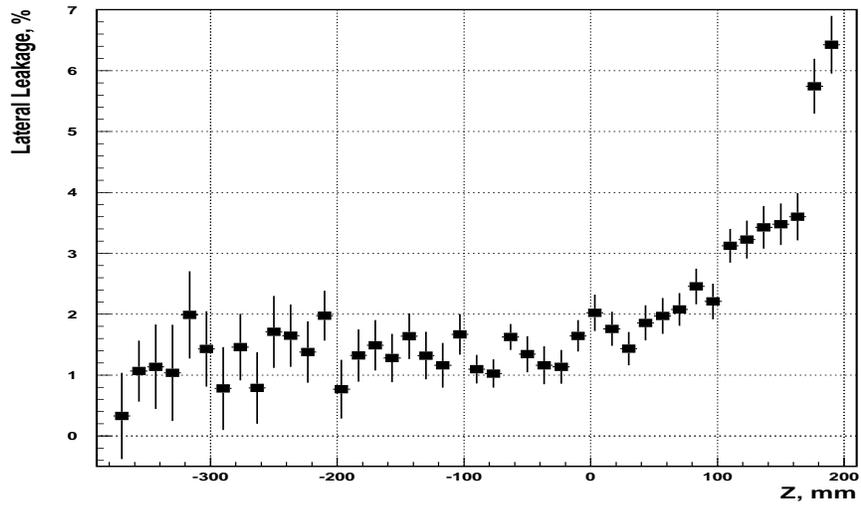,width=0.95\textwidth,height=0.4\textheight}} \\
        %\hline
        %\hline
        \mbox{\epsfig{figure=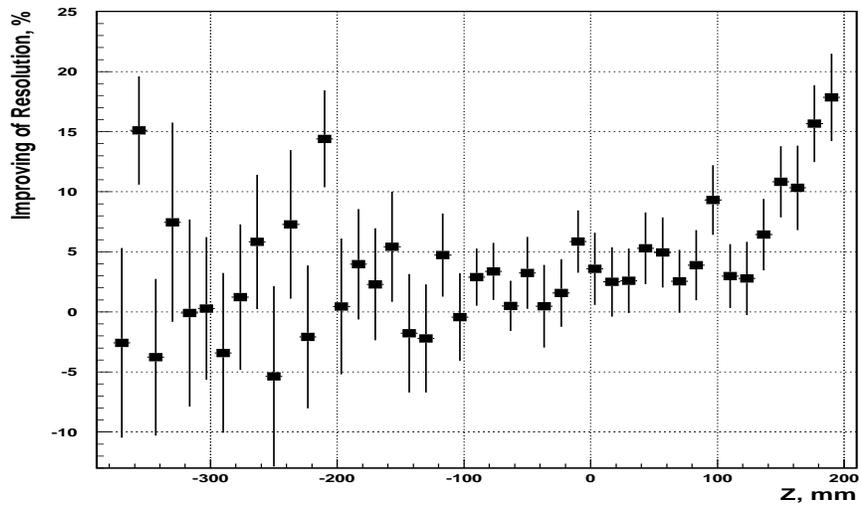,width=0.95\textwidth,height=0.4\textheight}} \\
        %\hline
        \end{tabular}
     \end{center}
      \caption{
        The lateral leakage (top) and the energy resolution improving
        (bottom) for the events sample with lateral leakage
        as a function of $Z$ coordinate.
       \label{fig:f12}}
\end{figure*}

\begin{figure*}[tbph]
     \begin{center}
        \begin{tabular}{c}
        %\hline
        \mbox{\epsfig{figure=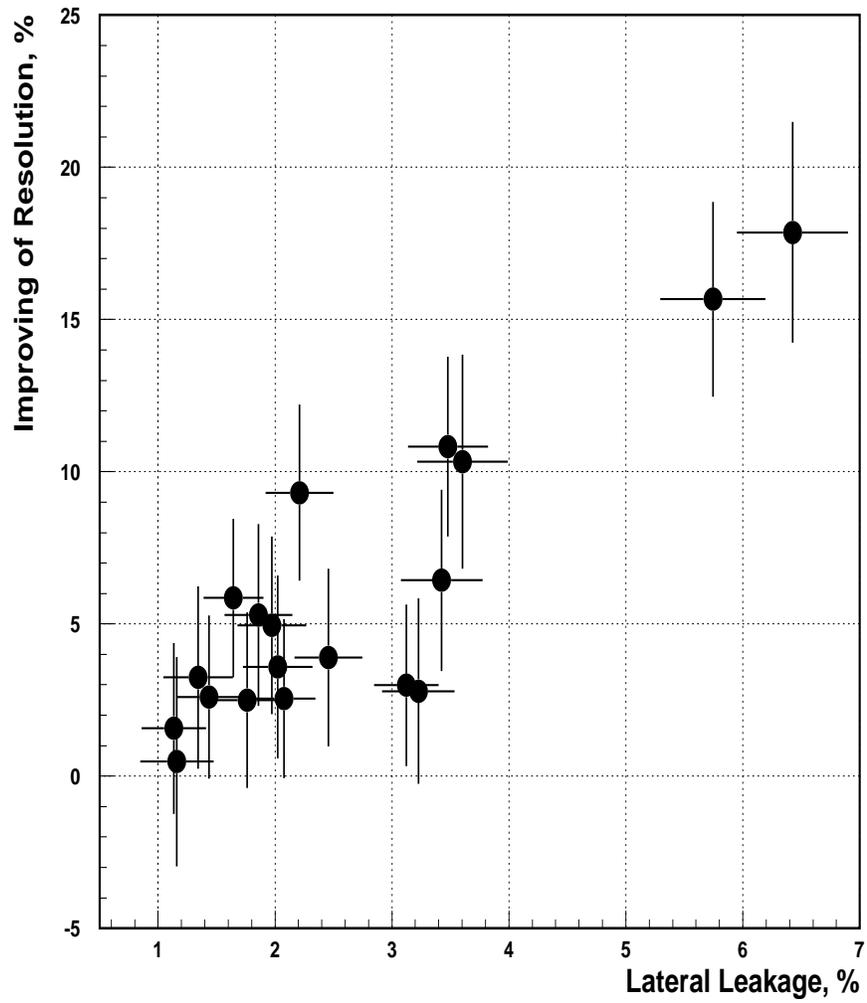,width=0.95\textwidth,height=0.8\textheight}} 
        \\
        %\hline
        \end{tabular}
     \end{center}
      \caption{
        The energy resolution improving for 
        the events sample with lateral leakage as a function of 
        lateral leakage for $Z > - 5\ cm$.
       \label{fig:f13}}
\end{figure*}

\end{document}